\documentclass{JHEP3} 
\preprint{ITEP-TH-34/08}

\JHEPspecialurl{http://jhep.sissa.it/JOURNAL/JHEP3.tar.gz}

\usepackage{epsfig,multicol,bbm}

\newcommand\fverb{\setbox\fverbbox=\hbox\bgroup\verb}
\newcommand\fverbdo{\egroup\medskip\noindent%
			\fbox{\unhbox\fverbbox}\ }
\newcommand\fverbit{\egroup\item[\fbox{\unhbox\fverbbox}]}
\newbox\fverbbox

\author{A.~V.~Zayakin\\Institute for
Theoretical and Experimental Physics\\ 117259, Moscow, B.
Cheremushkinskaya 25, Russia and\\ \\
Ludwig-Maximilians-Universit\"at M\"unchen,\\
Maier-Leibnitz-Labor\\
Am Coulombwall 1, D-85748, Garching, M\"unchen, Germany\\
	E-mail: \email{zayakin@itep.ru}}

\received{\today} 		
\accepted{\today}		


\usepackage{amsmath,amssymb,graphicx}
\usepackage{comment}
\newcommand{\beq}{\begin{equation}}

\newcommand{\eeq}{\end{equation}}
\newcommand{\myref}[1]{~{(\ref{#1})}}
\newcommand{\mycite}[1]{~{\cite{#1}}}
\newcommand{\myfigref}[1]{~{Fig.~(\ref{#1})}}

\def \be  {\begin{equation}}
\def \ee  {\end{equation}}
\def \ba  {\begin{eqnarray}}
\def \ea  {\end{eqnarray}}
\def \l   {\left}
\def \r   {\right}

\newcommand{\sym}{$\mathcal{N}=4$ SYM~}
\newcommand{\ads}{$AdS_5$~}

\title{QCD Vacuum Properties in a Magnetic Field from AdS/CFT:\\ Chiral Condensate and Goldstone Mass}

\abstract{Chiral condensate and $\eta^\prime$ meson mass spectrum are studied under the influence of an external Abelian magnetic field. We work within the D3/D7 Karch---Katz model of flavoured AdS/CFT with supersymmetry broken by the Constable---Myers deformation of the metric. It is shown that this setting yields an analytic (quadratic) dependence of condensate on field, typical for the Nambu---Jona-Lasinio model, rather than the non-analytic (linear in field) result, typical for chiral perturbation theory in the exact chiral limit. We conjecture that the analytic (quadratic) result might be put into correspondence with the leading-order in the $1/N_c$ decomposition for the condensate. This leading order in the $1/N_c$ approximation has not yet been derived from the chiral perturbation theory.  Thus the dual model yields the quadratic field dependence of the condensate, which is beyond the range of feasibility of chiral perturbation theory.}
\begin{document}

\section*{Introduction}
The behaviour of QCD vacuum in strong electromagnetic fields has recently attracted a great deal of attention (e.g.\mycite{Kabat:2002er,Miransky:2002rp}), reinvigorating the subject
which had been started by\cite{Ioffe:1983ju}. Lattice \hbox{simulations}\mycite{Cea:2006gu, Shintani:2006xr}, Simonov's string model\mycite{Kruglov:1997dy}  are just a few of the recent studies of QCD vacuum in external fields to be mentioned here. In this work we try to describe the behaviour of mesonic spectra and condensate from the perspective of duality. This article is organized as follows. In the following Section \ref{motivation} treatment of mesonic masses' and condensates' in external fields is reviewed. It is explained why traditional field-theoretical approaches, are still demanding a non-perturbative insight, possibly coming from the realm of dual models. Then in Section \ref{dual} a short description of the specific dual model is given, which we are going to apply. In the subsequent Section \ref{chircond} the numerical calculations are presented. We conclude in \ref{concl}.

\section{Motivation\label{motivation}}

The QCD vacuum is quantitatively described by its chiral condensate, gluonic condensate, pion decay constant and some other physical quantities. Below we shall revisit the properties of some of these objects in the electromagnetic background.

\subsection{Condensate}
QCD chiral condensate $\langle \bar{q}q\rangle$ is the order parameter of chiral symmetry breaking.
Important ideas of chiral symmetry breaking catalysis were being developed by Gusynin, Miransky and Shovkovy. In\mycite{Miransky:2002rp,Gusynin:1994re,
Gusynin:1994xp} they study enhancement of chiral  condensates in $2+1$ and $3+1$-dimensional Nambu---Jona-Lasinio-like models.

The issue of condensates in an external magnetic field was resolved by \hbox{Schramm}, Mueller and Schramm~\cite{Schramm:1991ex}, and by Smilga and Shushpanov in\mycite{Shushpanov:1998ms}. For small magnetic fields $H$, and in an exact chiral limit
 \beq\label{c-limit}
 \langle \bar{q}q\rangle_H=\langle \bar{q}q\rangle_0\l(1+\frac{eH\ln 2}{16\pi^2f_\pi^2}\r).
 \eeq Note that the linear term in $H$ has a $\frac{1}{N_c}$ factor, for $f_\pi^2\sim N_c$
 A second-loop correction to this result was calculated by Shushpanov and Agasian\mycite{Agasian:1999vh}.
 It is instructive to compare this low-energy QCD computation with a Nambu---Jona-Lasinio model computation made by Klevansky and Lemmer~\cite{Klevansky:1989vi}
 \beq
 \langle \bar{q}q\rangle_H=\langle \bar{q}q\rangle_0\l( 1+c\frac{e^2H^2}{(\langle \bar{q}q\rangle_0)^{4/3}}\r),
 \eeq
 where $c$ is some model-dependent coefficient.
The linear dependence\myref{c-limit} by Smilga and Shushpanov is non-analytic (has a square-root type cut) in terms of the invariants of the external field, i.e. is organized as $\sim\sqrt{F^2}$. This might seem to be inconsistent from a first view. However, this non-analyticity is of vital importance. It means there are no other massive parameters in the low-energy domain, where the chiral perturbation theory is valid. The non-analyticity of\myref{c-limit} is a direct signature of $\pi$-meson being a Goldstone particle. If chiral limit is violated, the dependence will be analytic. One must work here in the exact chiral limit, for otherwise all other massive hadronic states in the vacuum energy loops must be taken into account.  Nambu---Jona-Lasinio model is at the same time seen to be deficient to describe full QCD, as it does not reproduce the correct non-analytic behaviour of the condensate, representing the Goldstone particles.

Chiral condensates in arbitrary electromagnetic fields were calculated by Cohen, McGady and Werbos in~\cite{Cohen:2007bt}. They have obtained expressions for electric, magnetic, and arbitrary configuration of constant fields. Their results are basically obtained in the same Heisenberg-Euler technique type as those of Smilga and Shushpanov, and perfectly  reproduce the latter as a particular case.

\subsection{Limitations of traditional approaches}
The above chiral perturbation theory results have the status of exact low-energy theorems. However, they have their domain of applicability, as explained in the review paper by Ioffe\mycite{Ioffe:2002ee}. The one-loop result has been reminded above. This means there will be next-order loop corrections in chiral perturbation theory to this value. Chiral perturbation theory has also limitations due to the fact that quarks' and gluons' degrees of freedom are fully absent in it. Therefore, other models have to be considered to be compared with chiral perturbation theory estimates.

A class of modern (supersymmetric and non-supersymmetric) QCD models, which would be natural to test for the behaviour of condensates and meson masses, are AdS/CFT models with flavours. For a review on AdS/CFT with flavours or in non-supersymmetric backgrounds see\mycite{Aharony:2002up,Mateos:2007ay}. They are generally constructed basing on Maldacena's conjecture\mycite{Maldacena:1997re} on equivalence of the IIB type closed string theory in the bulk of \ads$\times S^5$ and \sym theory in the four-dimensional flat spacetime. For a review of Maldacena conjecture in general, see
\mycite{Aharony:1999ti,D'Hoker:2002aw}.
QCD is, of course, a non-supersymmetric and a non-conformal theory, so various symmetry breaking techniques are applied to make the model resemble the reality. The two most common approaches are: ``bottom-up'' approach, and ``top-down'' approach. A ``bottom-up'' construction, see e.g.\mycite{Erlich:2005qh}
is usually constructed with a 5-dimensional action, which one has to ``guess'', so that it fits as many QCD results as possible. On the other hand, the ``top-down'' approach (e.g. \mycite{Karch:2002sh}) is constructed starting with a 10-dimensional geometrical setting, in which special elements are supposed to reproduce the dynamics of QCD degrees of freedom.

Flavoured AdS/CFT correspondence in an external magnetic field was studied by Filev et al. in\mycite{Filev:2007gb}. They produce a spectrum of mesons from pure-AdS background, which satisfies the Gell-Mann---Oakes---Renner relation. In\mycite{Albash:2007bk} thermodynamic  properties of the gauge theory in a magnetic field have been studied in the same framework. Properties of the theory in an electric field were obtained in\mycite{Albash:2007bq} by the same method.

AdS/CFT with flavours in external fields and at finite temperatures have also been studied in\mycite{Erdmenger:2007bn}. The authors calculate a number of external-field-dependent properties for a supersymmetric background, such as meson masses in electric and magnetic fields. Sakai---Sugimoto model in external fields was studied in~\cite{Johnson:2008vn}. It has been concluded that Sakai---Sugimoto model is consistent with the picture of magnetic catalysis of chiral symmetry breaking. Phase transitions in Sakai---Sugimoto models due to switching on of electric and magnetic fields were discussed in\mycite{Bergman:2008sg}. Pair production in an electric field in Sakai---Sugimoto model was studied in\mycite{Kim:2008zn}. When the present paper was being completed, two works on holographic QCD at finite temperature, magnetic field and chemical potential appeared on the same day~\mycite{Thompson:2008qw,Bergman:2008qv}.
  This is the evidence for the great interest to the different aspects of AdS/CFT in external Abelian fields that is present nowadays.

\section{D3/D7 model in Constable---Myers background with a Kalb---Ramond field\label{dual}}
In this short letter a very simple model is discussed, which  features many of the basic QCD characteristics: confinement, conformal symmetry breaking and  spontaneous chiral symmetry breaking.  We put it into a magnetic field and observe the behaviour of condensates and mass spectra.

Below we follow what is known as Karch-Katz model with Constable---Myers deformation. We rely in this passage essentially (sometimes literally) on\mycite{Babington:2003vm}. This geometry conjecturally describes a \sym broken by non-zero expectation values for all $SO(6)$ singlet operators. It also inherits from pure Karch---Katz model a pack of $D7$ probe branes, which do not affect the metrics. This requires us to work in the quenched approximation $N_f\ll N_c$. The Constable ---Myers metric is organized as
\beq
ds^2=H^{-\frac{1}{2}}
\l(1+\frac{2b^4}{r^4}\r)^{\frac{\delta}{4}}dx^2+
H^{\frac{1}{2}}\l(1+
\frac{2b^4}{r^4}\r)^{\frac{2-\delta}{4}}
\frac{r^2}{\l(1+
\frac{2b^4}{r^4}\r)^{\frac{1}{2}}}
\l[\frac{r^6}{(r^4+b^4)^2}dr^2 +d\Omega^2\r],
\eeq
where
\beq
H=\l(1+\frac{2b^4}{r^4}\r)^\delta -1.
\eeq
This form of the metric makes it easy to see that it behaves asymptotically at $r\to\infty$ as pure AdS, but differs from it near the singularity.

The Constable---Myers solution requires a non-trivial dilaton as well
\beq
e^{2\phi}=e^{2\phi_0}
\l( 1+\frac{2b^4}{r^4}\r)^{\Delta}
\eeq
and a $C_4$ form field
\beq
C_{(4)}=-\frac{1}{4}H^{-1} dx_0\wedge dx_1\wedge dx_2\wedge dx_3,
\eeq
with conditions imposed upon deformation parameters
\beq
\begin{array}{l}
\Delta^2+\delta^2=10,\\
\delta=\frac{1}{2b^4}.
\end{array}
\eeq
For a more convenient embedding of the $D7$ brane, a coordinate transformation is performed, which will explicitly separate the 4-dimensional and 6-dimensional hyperplanes:
\beq
ds^2=H^{-\frac{1}{2}}
\l(\frac{w^4+b^4}{w^4-b^4}\r)^{\frac{\delta}{4}}dx^2+
H^{\frac{1}{2}}
\l(\frac{w^4+b^4}{w^4-b^4}\r)^{\frac{2-\delta}{4}}
\frac{w^4-b^4}{w^4}\sum_{i=1}^6dw_i^2,
\eeq
where now
\beq
H=\l(\frac{w^4+b^4}{w^4-b^4}\r)^\delta-1
\eeq
and the dilaton is
\beq
e^{2\phi}=e^{2\phi_0}
\l(\frac{w^4+b^4}{w^4-b^4}\r)^{\Delta}
\eeq
We have already learned about the equivalence of Kalb---Ramond field in the bulk and the Maxwell field on the brane. The $D7$ brane does not change the metric in the quenched approximation. The dynamics of the brane is described by a Dirac---Born---Infeld action
\beq
S_{D7}=\mu_7\int d^{8} \xi \sqrt{\det_{\alpha,\beta}\l(2\pi B_{\alpha\beta}+2\pi\alpha^\prime F_{\alpha\beta}+
g_{\mu\nu}\frac{\partial X^\mu}{\partial \xi^\alpha}\frac{\partial X^\nu}{\partial \xi^\beta }\r)}+
\int d^{8} \xi C_4\wedge F \wedge B
\eeq
Here $B_{\mu\nu}$ is the Kalb---Ramond field, defined in the bulk, which is projected to the brane as $B_{\alpha\beta}$ and $F_{\alpha\beta}$ is the usual Maxwell field on the brane. A constant field $F_{23}=-F_{32}=B$ is chosen, all other field components being zero.
The Chern---Simons term does not influence classical dynamics. It may give a contribution into the oscillations describing mesonic masses. Further the embedding geometry and gauge condition are specified. The D7 brane runs through the directions of coordinates $x_0,x_1,x_2,x_3,w_1,w_2,w_3,w_4$. These coordinates are respectively $\xi_1\dots \xi_8$ internal coordinates of the brane world-volume. It doesn't run through the remaining $w_5,w_6$. The latter coordinates are embedding coordinates of the brane into the targetspace. They are functions of $\xi_i$. Solutions in the form $w_5=w(\rho),w_6=0$ will be sought, where
\beq
\rho=\sqrt{w_1^2+w_2^2+w_3^2+w_4^2}.
\eeq
With such an Ansatz and in the metrics given above, the DBI action is organized as
\beq
S=-\mu_7\int d^8 \xi G(\rho,w)\sqrt{1+w^{\prime 2}(\rho)}
\sqrt{1+B^2/g_{11}^2},
\eeq
where \beq
G(\rho, w)=\rho^3 \frac{\l((\rho^2+w^2)^2+b^4\r)
\l((\rho^2+w^2)^2-b^4\r)}{(\rho^2+w^2)^4}e^{2\phi}\eeq
The equations of motion will look like
\beq
\frac{d}{d\rho} \l(\frac{G w^\prime}{\sqrt{1+w^{\prime 2}}}
\sqrt{1+B^2/g_{11}^2}\r)-\sqrt{1+w^{\prime 2}}\frac{d}{d w}\l(G \sqrt{1+B^2/g_{11}^2}\r)=0
\eeq
They are solved them numerically in the next section. First quark masses and condensates are extracted and fitted with appropriate interpolation functions. Then small oscillations around the classical solutions are studied. The spectra of these oscillations are identified with meson masses according to known rules\mycite{Erdmenger:2007cm} of AdS/CFT correspondence.

\section{Condensate and spectra\label{chircond}}
\subsection{Condensate}
The standard lore is: one must search for physical solutions of these non-linear
\FIGURE{
\includegraphics[height = 10cm, width=21cm]{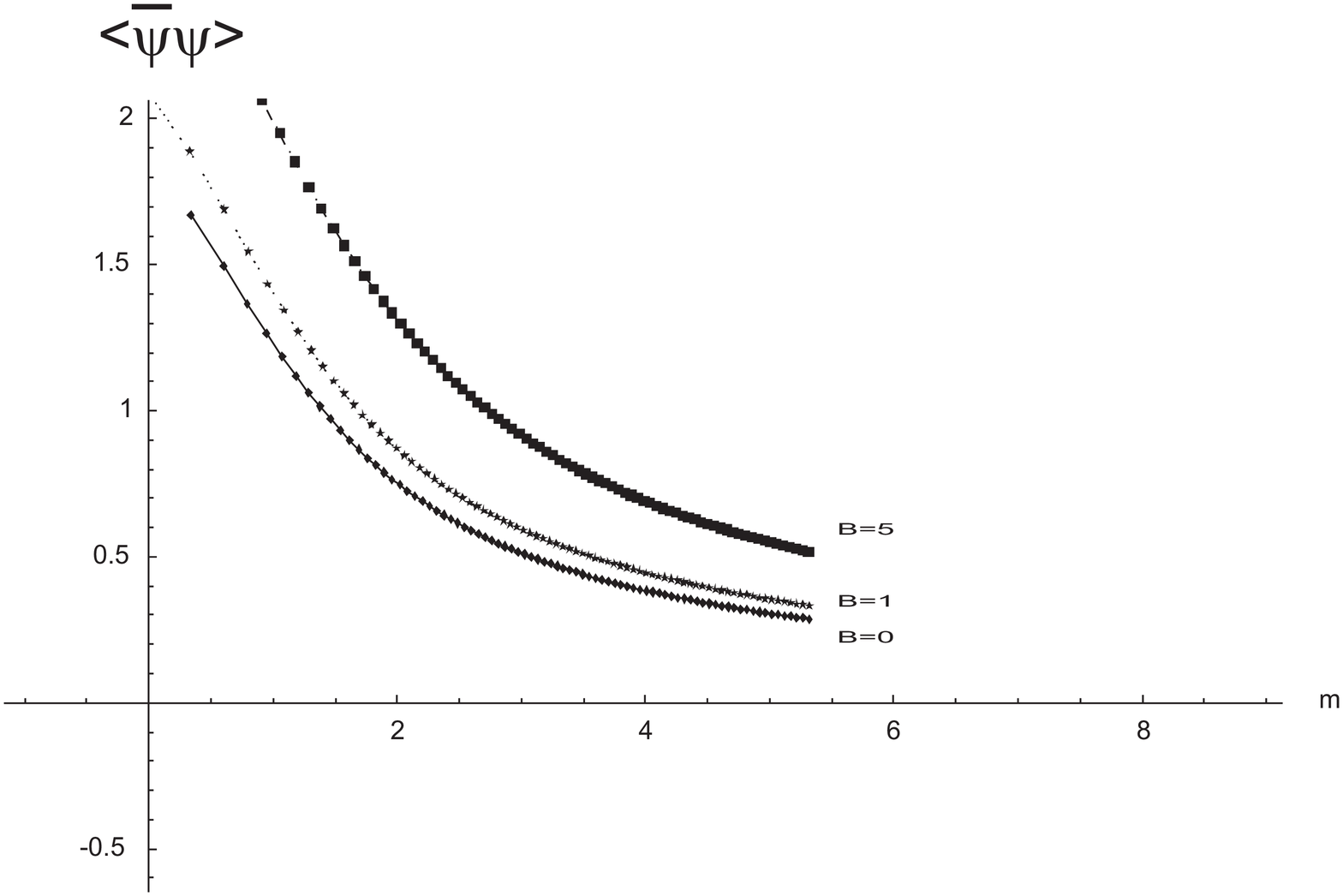}
\caption{Dependence of condensate on magnetic field and mass.} \label{chiral}
}
\FIGURE{
\includegraphics[height = 7cm, width=10cm]{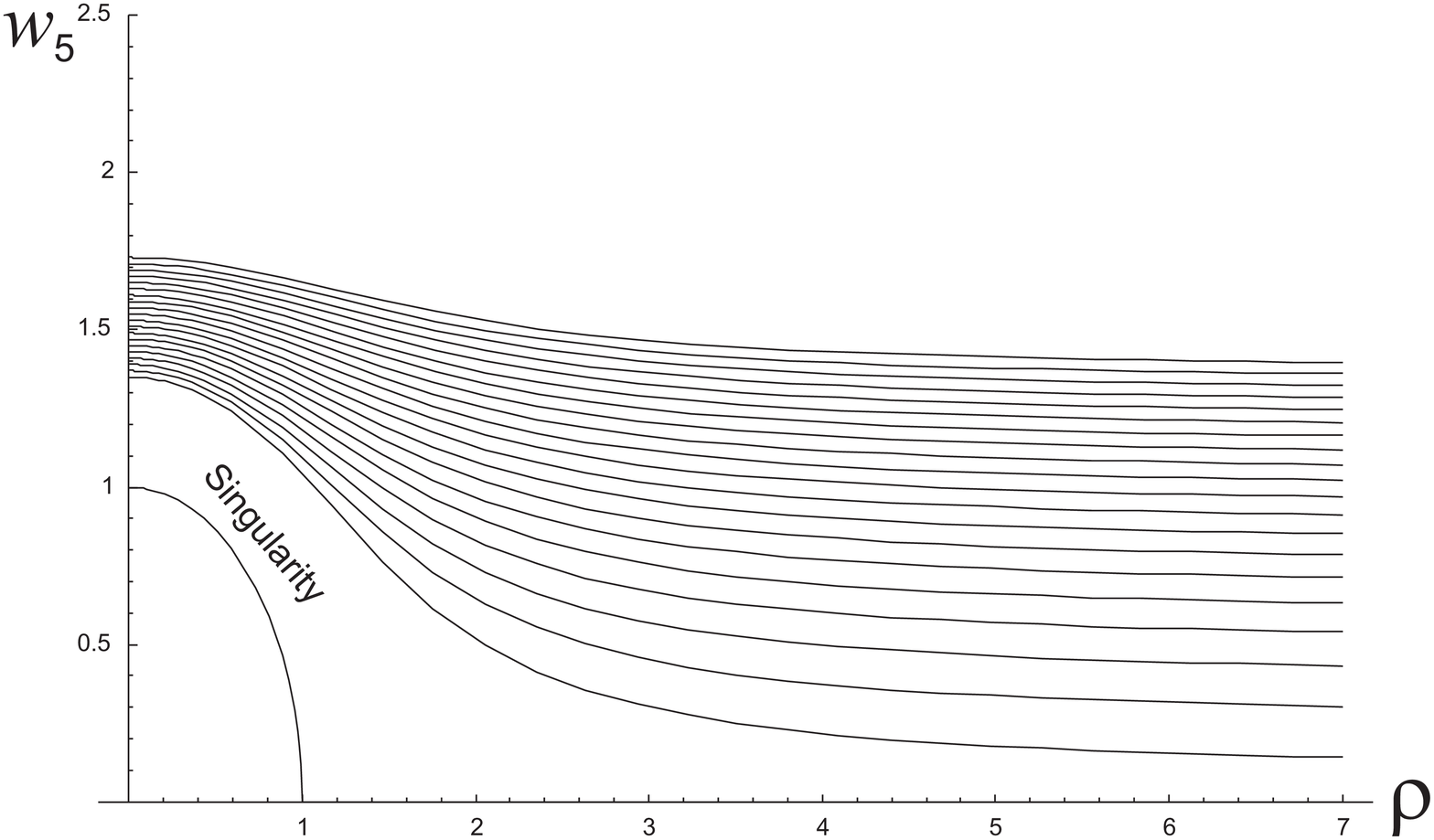}
\caption{Different embeddings of the spectator D7 brane into Constable---Myers background.} \label{Solutions}
}
\FIGURE{
\includegraphics[height = 10cm, width=16cm]{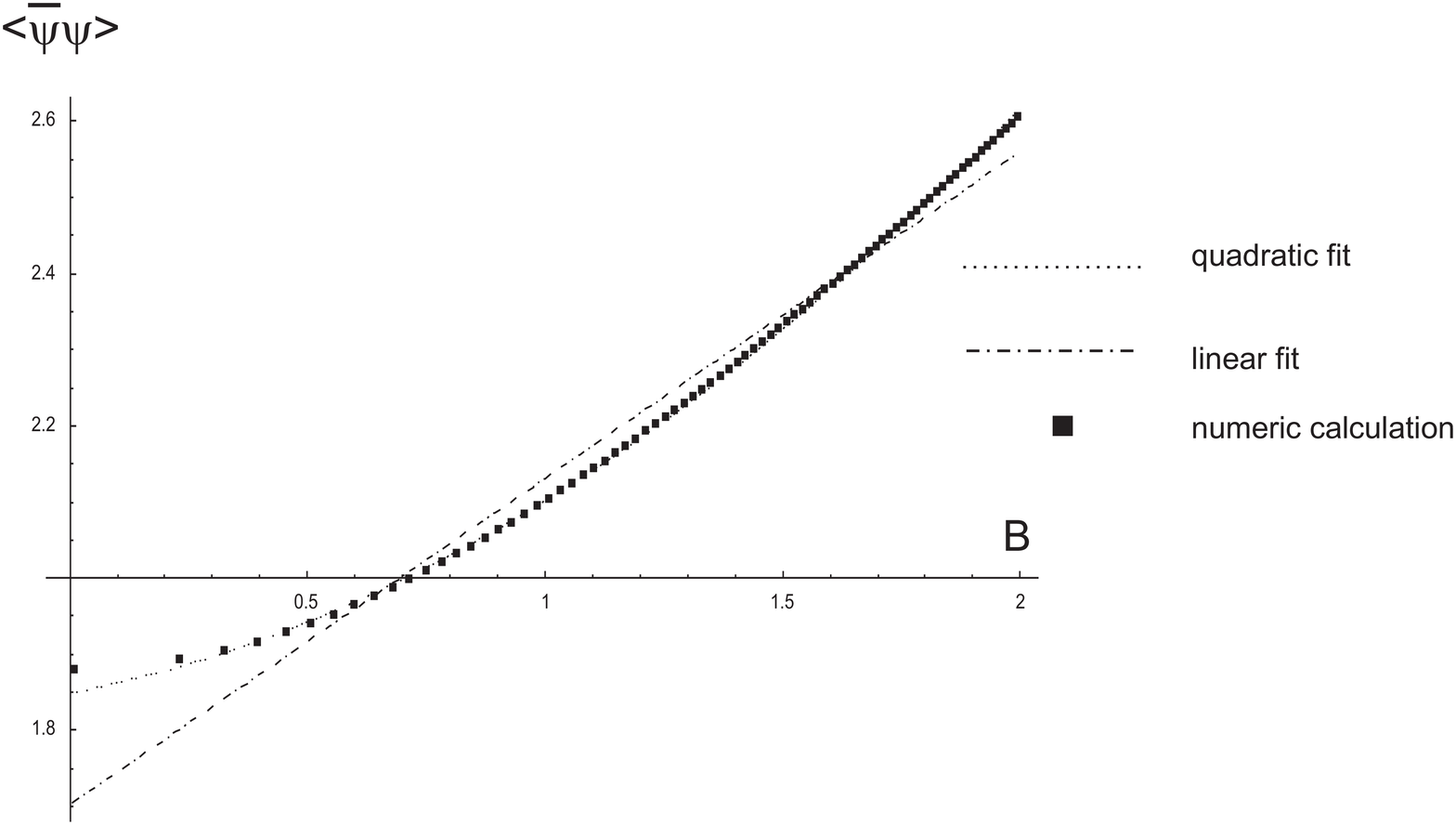}
\caption{Magnetic catalysis of chiral symmetry breaking in Karch---Katz model with Constable---Myers deformation, exact chiral limit $m=0$.} \label{catalysis}
}
\FIGURE{
\includegraphics[height = 6cm, width=15cm]{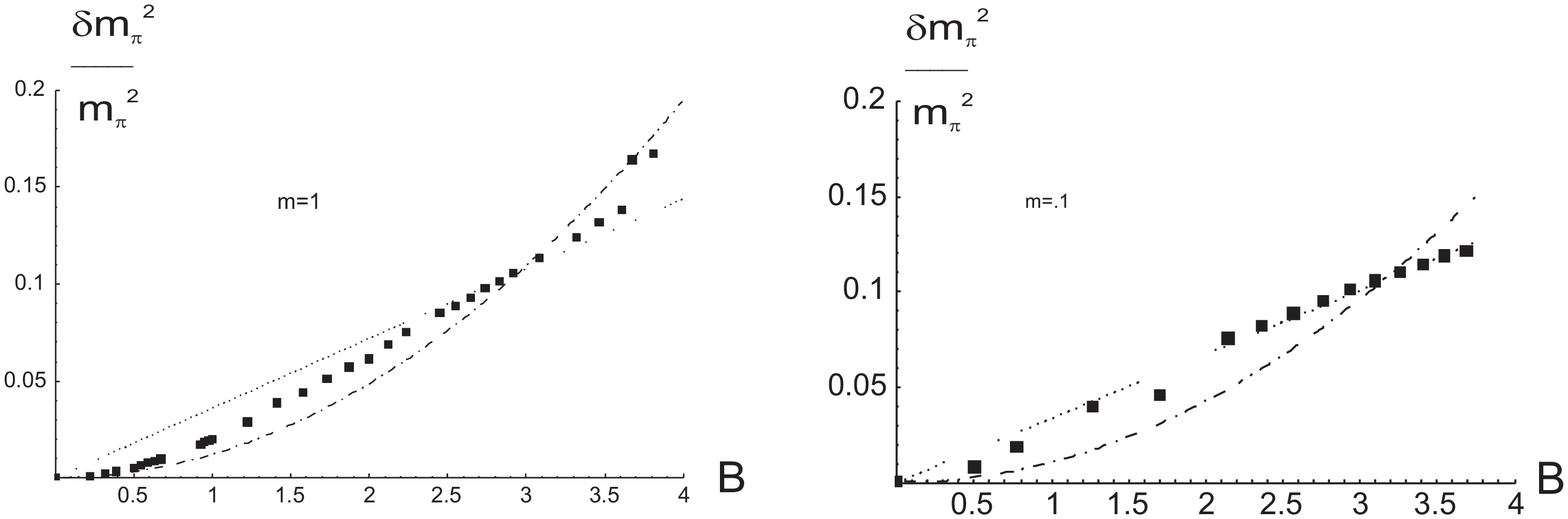}
\caption{Mass spectra of $m_\pi^2$ as functions of $B$. Thin lines on the left ($m=1$) and right ($m=0.1$) plots show interpolation $\delta m_\pi^2=\alpha B$ and $\delta m_\pi^2=\alpha B^2$. One can see that neither of these interpolations is satisfactory.} \label{massspectra}
}second-order differential equations, which have the following asymptotics in the infinity:
\beq
w(\rho)=m+\frac{c}{\rho^2}.
\eeq
Then the parameters $m$ and $c$ correspond to quark mass and chiral condensate:
\beq
\begin{array}{l}
m_q=\frac{m}{2\pi\alpha^\prime},\\
\langle\bar{q}q
\rangle=\frac{c}{(2\pi\alpha^\prime)^3},
\end{array}
\eeq
where $\alpha^\prime$ is string tension parameter. Contrary to the physical solutions, unphysical ones are those ending in the singularity of Constable---Myers metrics, or going to infinity at $\rho\to 0$. The singularity is marked by an ellipse denoted ``singularity'' in\myfigref{Solutions}. Physical solutions can be defined by boundary condition $w^\prime(0)=0$.
It happens that the generic  solutions are unphysical ones.

To obtain physical solutions, one imposes
\beq\begin{array}{l} w'(0)=0,\\ w(0)=w_0=const.\end{array}\eeq
For each value of $w_0$ above some value $w_{min}\approx 1.34$ this will yield a curve\myfigref{Solutions}, asymptotic behaviour of which will reveal some definite $m$ and $c$. This allows one to build the dependence of condensate on quark mass\myfigref{chiral}. Doing the same thing with different values of magnetic field $B$ one gets a shifted curve. It's a subtlety of this method that in order to understand how the condensate shifts in the field, one must take a section of\myfigref{chiral} at a constant $m$ rather than follow some definite point in the plot. The resulting dependence of condensate on the field is shown in\myfigref{catalysis}. It is noted here that the results for $B=0$ coincide with those calculated in the same background in\mycite{Babington:2003vm}.

We analyze the ``experimental'' dependence. It is natural to expect either linear (as in true QCD) or quadratic (as in NJL) condensate growth with the field. In our case, approximation with a quadratic polynomial comes out to be quite effective. In the picture\myfigref{catalysis} one can see the comparison between the linear and quadratic approximations, and judge in favour of the latter.

This quadratic dependence on field value corresponds very nicely to the picture of magnetic catalysis of chiral symmetry breaking in~\cite{Gusynin:1994xp,Miransky:2002rp} in NJL models. On the other hand, it does not correspond to linear condensate shift, predicted by the low-energy QCD effective action by Smilga and Shushpanov\mycite{Shushpanov:1998ms}. This phenomenon may be given a nice qualitative explanation. Condensate expression\myref{c-limit} is a part of the series in powers of $\frac{1}{N_c}$, for $f_\pi\sim \sqrt{N_c}$. It starts with the $\frac{1}{N_c}$ term. There may be a term, dependent on field, and containing $\frac{1}{N_c}$ in the zeroth power. To our best knowledge, such terms have not been reported in chiral perturbation theory. On the contrary, dual models restore the missing leading-order $\frac{1}{N_c}$ contribution.

\subsection{Meson spectra}
Small fluctuations of the classical solutions to the equations of motion govern mesonic spectra.
There are two types of these fluctuations: those corresponding to Goldstone (in the large $N_c$ limit) mesons $eta^\prime$, and those corresponding  to the non-Goldstone ones. The former are fluctuations of the angular coordinate in $Ox_8x_9$ plane, the latter are the fluctuations of the radial coordinate. The equations for Goldstone part of the spectra are\mycite{Babington:2003vm}:
\beq\begin{array}{l}\label{fluct}
\frac{d}{d\rho}\l[\frac{G \sqrt{1+B^2/g_{11}^2}}{\sqrt{1+w^{\prime 2}}}
\partial_\rho f(\rho)\r]+M^2 \frac{G \sqrt{1+B^2/g_{11}^2}}{\sqrt{1+w^{\prime 2}}}H \l(\frac{(\rho^2+w^2)^2+b^4}{(\rho^2+w^2)^2-b^4}
\r)^{\frac{1-\delta}{2}}
\frac{(\rho^2+w^2)^2-b^4}{(\rho^2+w^2)^2}f(\rho)\\
-\sqrt{1+w^{\prime 2}}\sqrt{1+B^2/g_{11}^2}
\frac{4b^4\rho^3}{(\rho^2+w^2)^5}
\l(\frac{(\rho^2+w^2)^2+b^4}{(\rho^2+w^2)^2-b^4}
\r)^{\frac{\Delta}{2}}\l(2b^4-
\Delta(\rho^2+w^2)^2\r)f(\rho)=0.
\end{array}
\eeq
This is a Sturm---Liouville eigenvalue problem on function $f(\rho)$, which must be solved with the following boundary conditions:
\beq\left\{\begin{array}{l}
f^\prime(0)=0\\
f(\rho)|_{\rho\to\infty}\to\frac{1}{\rho^2}.
\end{array}\right.
\eeq
The function $w$ in the equation\myref{fluct} must be taken from the previous section. These  equations are solved in {\it Mathematica} environment by shooting method. One starts with solutions behaving like $\frac{1}{\rho^2}$ in the infinity, and step-by-step find the value of $M$, which yields the desired behaviour at the origin. Again, the resulting regular solution is never reached, but it can be approximated as a separatrix of solutions singular at origins, to any desired accuracy. In our calculations we have obtained $M^2$ with 4 decimal digits.
The results for Goldstones are  shown in\myfigref{massspectra}, where
\beq
\delta m_\pi^2=m_\pi^2(B)-m_\pi^2(0).
\eeq
 We have tried, basing on previous experience both on field theory side and gravity side, to approximate the field dependence of $\delta m_\pi^2$ by either linear $\delta m_\pi^2\sim B$ or quadratic $\delta m_\pi^2\sim B^2$ dependence. However, our numerical analysis has shown that neither can be a valid approximation, even for small $B$, which is shown in\myfigref{massspectra}. Comparing this to the linear dependence for masses in the chiral limit of chiral perturbation theory and to the quadratic dependence given for pure AdS background in \cite{Erdmenger:2007bn}, we conclude that dynamics of our model might be away from both the predictions of chiral perturbation theory and pure AdS model.

\section{Conclusion\label{concl}}
A qualitative conclusion can be drawn upon analyzing the dependence of condensates on the field. We can see that the linear field dependence of QCD condensate from chiral perturbation theory is not reproduced at all. Instead, a quadratic dependence is retrieved. Our conjecture to explain this phenomenon is very simple. Chiral perturbation theory estimate, as given in the cited references, misses the leading-order in $\frac{1}{N_c}$. It starts with the next-to-leading order in $\frac{1}{N_c}$. On the other hand, duality might reproduce the leading-order effect.
Nevertheless, the search for a true dual model of QCD must still be in progress, for meson mass spectra cannot be easily given a qualitative explanation.  One of possible improvements of the model would be to take into account back-reaction effects. In our setting, the D7 brane was a probe brane, a self-consistent supergravity solution in a background of a stack of D3 branes and a D7 brane would be more complicated.

\section*{Acknowledgements}  I thank A.~Gorsky and D.~V.~Shirkov for their attention to this work, their comments and advice. I thank J.~Rafelski for posing the question to which this paper hopes to provide an answer. I am grateful to E.Akhmedov, A.~Andrianov, D.~Diakonov,  J.~Erdmenger, M.~Haack, P.~Koroteev, R.~Meyer, A.~Smilga, and A.~Vainshtein  for fruitful discussions.  I.Kirsch and R.Meyer have helped me enormously in understanding their earlier work and numeric implementation of the D3/D7 model. Thanks to O.~Andreev, N.~Agasian, I.~Shovkovy, I.~Shushpanov, V.I.~Zakharov for comments on their prior work.   I am also grateful to the friendly faculty and staff of LMU-Munich, and ITEP-Moscow. This work is supported in part by RFBR Grant 07-01-00526 and DFG Cluster of Excellence MAP, Munich.

\end{document}